\documentclass[12pt]{article}
\usepackage{latexsym}
\usepackage{amsmath}
\renewcommand{\epsilon}{\varepsilon}

\usepackage{amsmath,bm}
\usepackage{amssymb}
\usepackage{xcolor}
\usepackage{amsthm}
\usepackage[ansinew]{inputenc}
\usepackage{graphicx}
\usepackage{epsfig}
\usepackage{dsfont}
\usepackage{bbm}
\usepackage{url}
\usepackage{placeins}
\usepackage{subfigure}
 \usepackage{multirow}

 \usepackage[authoryear]{natbib}
\newtheorem{satz}{Theorem}[section]

\newtheorem{algorithm}[satz]{Algorithm}

\def\3{\ss}

\newcommand{\bea}{\begin{eqnarray*}}
	\newcommand{\eea}{\end{eqnarray*}}
\newcommand{\be}{\begin{eqnarray}}
\newcommand{\ee}{\end{eqnarray}}

\newcommand{\ba}{\begin{array}}
	\newcommand{\ea}{\end{array}}

\def\3{\ss}

\begin{document}
	
	\title{{\bf Efficient model-based Bioequivalence Testing}}

	\author{\small Kathrin M\"ollenhoff, Florence Loingeville, Julie Bertrand, Thu Thuy Nguyen, \\
		\small Satish Sharan, Guoying Sun, Stella Grosser, Liang Zhao, Lanyan Fang, \\
		\small France MentrÚ, Holger Dette \\ \\
	\small Ruhr-Universit\"at Bochum, Fakult\"at f\"ur Mathematik, 44780 Bochum, Germany, \\
	\small UniversitÚ de Paris, IAME INSERM, 75018 Paris, France,\\
    \small Faculty of Pharmacy, Univ. of Lille, EA 2694: Public health: \\
		\small Epidemiology and Healthcare quality, 59000 Lille, France,\\
    \small Division of Quantitative Methods and Modeling, Office of Research Standards, \\
		\small Office of Generic Drugs, Center for Drug Evaluation and Research,\\
    \small Food and Drug Administration, 10903 New Hampshire Ave Silver Spring MD 20993, USA,\\
    \small Office of Biostatistics, Office of Translational Sciences, Center for Drug Evaluation and Research,\\
    \small Food and Drug Administration, 10903 New Hampshire Ave Silver Spring MD 20993, USA\\
	}

	\pdfminorversion=4
	\maketitle
	
	\begin{abstract}

The classical approach to analyze pharmacokinetic (PK) data in bioequivalence studies aiming to compare two different formulations is to perform noncompartmental analysis (NCA) followed by two one-sided tests (TOST). In this regard the PK parameters $AUC$ and $C_{max}$ are obtained for both treatment groups and their geometric mean ratios are considered. According to current guidelines by the U.S. Food and Drug Administration and the European Medicines Agency the formulations are declared to be sufficiently similar if the $90\%$- confidence interval for these ratios falls between $0.8$ and $1.25 $.
As NCA  is not a reliable approach in case of sparse designs, a model-based alternative has already been proposed for the estimation of $AUC$ and $C_{max}$ using non-linear mixed effects models.
Here we propose another, more powerful test than the TOST and demonstrate its superiority through a simulation study  both for NCA and model-based approaches. For products with high variability on PK parameters, this method appears to have closer type I errors to the conventionally accepted significance level of $0.05$, suggesting its potential use in situations where conventional bioequivalence analysis is not applicable.

\end{abstract}


\parindent 0cm

\maketitle

\section{Introduction}
\label{sec1}
\def\theequation{1.\arabic{equation}}
\setcounter{equation}{0}

In  drug development  the comparison of two different formulations of the same drug is a frequently addressed issue. In this regard bioequivalence studies investigating the difference between two treatments are performed. According to current guidelines by the \cite{food2003guidance} and the \cite{ema} this question is commonly addressed by comparing the ratios of the geometric means of the pharmacokinetic (PK) parameters area under the curve ($AUC$) and the maximal concentration ($C_{\max}$) to a prespecified threshold. More precisely, bioequivalence is established if the boundaries of the $90\%$-confidence intervals for these ratios fall between $0.8$ and $1.25$ which is equivalent to performing two one-sided tests (TOST) proposed by  \cite{schuirmann1987}.
As the data are usually  log-transformed, we consider the  log-ratio (also defined as the treatment effect) and hence the commonly used equivalence margin is given by $\delta=\log(1.25)$.\\
When performing bioequivalence studies, the classical approach to analyze PK data is given by noncompartmental analysis (NCA), see for example \cite{gabrielsson2001}, followed by a linear mixed effect analysis of the AUC or Cmax.
The advantage of this approach is that it is very simple and comes without any further assumptions or knowledge of the data.
However, it requires a sufficiently large number of samples and subjects which cannot be provided in each trial.
As pointed out by  \cite{dubois2011}  and \cite{hu2004} the estimates obtained by NCA are  biased
if these conditions are not fulfilled.
Further, in numerous studies  a sufficiently large number of samples cannot be guaranteed. For instance, in pediatric research, ethical considerations lead to difficulties in the planning of studies which are therefore typically very small in size (for an example see \cite{mentre2001}).
But also in other areas where patients are especially frail, as for example in cancer research, these requirements are often not met and therefore methods for sparse designs are required.
In such situations the Nonlinear Mixed Effects Models (NLMEM)  have become very popular for  analyzing pharmacokinetic data  (see  \cite{sheiner1999}). NLMEM  turned out to be a promising alternative to the classical approach as the estimation of individual effects allows for incorporating variabilities, as the Between-subject-variability (BSV) and  the Within-subject-variability (WSV), for a detailed comparison see \cite{pentikis1996,combrink1997,panhard2005}.
Consequently the main advantage of the NLMEM  consists in the improved accuracy of the estimates in particular when dealing with sparse designs (see also \cite{hu2004}).\\
In order to assess bioequivalence between two products typically the two one-sided tests (TOST) proposed by  \cite{schuirmann1987} is performed, where two level $\alpha$-tests are combined for testing two seperate sub-hypotheses. This method is based on the Intersection-Union Principle (see \cite{berger1982})  and one  concludes bioequivalence if for both
one-sided tests the null hypotheses can be rejected. Due to its simplicity, this approach which is still recommended in the FDA guidelines has become very popular and is common practice nowadays (see for example \cite{bristol1993}, \cite{brown1997} and \cite{midha2009bioequivalence}  among many others).
However, it was demonstrated by \cite{phillips1990} and  \cite{tsai2014} that for a small number of individuals, high variability in the data or only few samples per patient this method is
rather  conservative and suffers from a lack of power.

The present paper addresses this problem.
Here we propose a new  model-based approach for the assessment of bioequivalence which turns out to have  always  more power than the corresponding  TOST.
The superiority  of the new approach  is  particularly visible in situations  with a large variability in the data in parallel designs.
The motivation of the new methodology is given by the uniformly
most powerful test  for  normally distributed data  with known variance, which can be found in
many text books on mathematical statistics
(see for example, \cite{lehmann2006}, or   \cite{wellek2010testing}).
We argue that the superiority  of this methodology for NCA also carries over to model-based inference
for reasonable large sample sizes and demonstrate  this fact  by means of a simulation study.

This paper is organized as follows. In Section \ref{sec2} we present the  classical problem of bioequivalence
and review two tests for this problem, including the commonly used TOST for NCA-based inference. 
 In Section \ref{sec4} we introduce the  NLMEM, then we present the model based TOST as first introduced by \cite{panhard2005} and \cite{dubois2011} and after that the new model based approach.
Subsequently, these tests are compared by means of  a  simulation study in Section \ref{sec5} with NCA-based tests both for parallel and cross-over designs varying BSV and WSV. In particular
we demonstrate  that the new approach (model and NCA-based) usually yields larger power than methodology
based on the TOST. 
Some theoretical arguments for  these finding can be found  in the Appendix, where we review properties of both
methods  in the  problem of comparing the means from two normal distributions with known variance. This
scenario corresponds to some kind of asymptotic regime for the problems considered in practice, if the sample sizes are
reasonably large. \\
Summarizing, the new  approach introduced in the present  paper improves the commonly used  TOST for bioequivalence testing based on NCA  or model-based inference.
It has never lower power than this test, but substantially larger power in scenarios with a large variability in parallel designs.


\section{Review of bioequivalence tests}
\label{sec2}
\def\theequation{2.\arabic{equation}}
\setcounter{equation}{0}

In this section we will briefly review a commonly used  approach for bioequivalence testing, which is based on the well-known two one-sided test (TOST)
introduced  by  \cite{schuirmann1987}.
 We further present another more powerful method for testing bioequivalence (see for example \cite{wellek2010testing}) which gives the motivation for the newly developed model-based test in Section \ref{sec43}.
 For the sake of simplicity, both methods are described here in the case of a two groups parallel design, but can be applied to crossover design, more standard in BE. 

In  a bioavailability/bioequivalence study  a test (T)  and a reference product (R) are administered and it is investigated
 whether the two formulations of the  drug  have similar properties with  respect to average bioavailabilty in the population.
 Exposure, in this context, is usually  characterized by blood concentration profile variables and summarized by the area under the time concentration curve (AUC) and
 the maximum concentration ($C_{\max}$). More precisely, let   $\mu_{T}$ and $\mu_{R}$ denote  the average means
 of the test and  reference product for $\log{AUC}$ or $\log C_{\max}$,
 then the common testing problem  in bioequivalence is defined by
 the hypotheses
\be\label{hypotheses}
H_0: \left|\mu_T-\mu_R\right|\geq \delta \text{ vs. } H_1:\left|\mu_T-\mu_R\right|< \delta,
\ee
where $\delta$ is a given threshold.  For example,   according to the $80/125$-rule considered in the guidelines by \cite{ema} and \cite{food2003guidance}
the threshold $\delta$ is given by $\delta=\log{(1.25)}$.

For the problem of testing for PK bioequivalence the metrics of interest are given by $AUC$ and $C_{\max}$, which means that we consider
\begin{eqnarray}\label{hyp_BE}
	\beta_{AUC}^T &:=& \mu_T-\mu_R=\log{AUC_T}-\log{AUC_R}\nonumber\\
	\beta_{C_{\max} }^T &:=&
	\mu_T-\mu_R=\log{C_{{\rm \max},T}}-\log{C_{{\rm \max},R}}
\end{eqnarray}
in \eqref{hypotheses},
where $\beta_{AUC}^T$ and $\beta_{C_{\max} }^T$ are the treatment effects on $AUC$ and $C_{\max} $ respectively.

\subsection{The two one-sided  Tests (TOST)}
\label{sec21}
We consider the following  sub-hypotheses of $H_0$ as described in \eqref{hypotheses} given by
\be\label{sub_hypotheses}
H_{0,-\delta}: \mu_T-\mu_R\leq -\delta \text{ and } H_{0,\delta}: \mu_T-\mu_R\geq \delta.
\ee
The idea of the TOST consists in testing each of these hypotheses separately by a one-sided test. The global null hypothesis $H_0$ in \eqref{hypotheses} is rejected with a type I error $\alpha$ if both one-sided hypotheses are rejected with a type I error $\alpha$.
To be precise let   $X_{T,1}, \ldots X_{T,N_{T}} $ and $X_{R,1}, \ldots X_{R,N_{R}} $ denote the samples from the test (T)  and a reference product (R) respectively
and denote by    $\bar{X}_k=\tfrac{1}{N_k}\sum_{i=1}^{N_k}X_{k,i}$  ($k=R,T$) the mean measured endpoints (over all individuals for the two treatments).
Under the assumption that the random variables $\{ X_{k,i} ~:~i=1,\ldots N_k, ~k=R,T \} $ are independent and
normally distributed with a common (but unknown) variance $\sigma^2$, that is  $X_{R,i} \sim {\cal N} (\mu_{R}, \sigma^{2}) $; $i=1 \ldots , N_{R}$~,~~$X_{T,i} \sim {\cal N} (\mu_{T}, \sigma^{2}) $; $i=1 \ldots , N_{T}$
we have  for the corresponding means
\begin{equation}\label{normality}\bar{X}_R\sim\mathcal{N}(\mu_{R},\tfrac{\sigma^2}{N_R})\text{ and } \bar{X}_T\sim\mathcal{N}(\mu_{T},\tfrac{\sigma^2}{N_T}) .\end{equation}
In applications $X_{R,i}$ and $X_{T,i}$ usually represent $AUC_k$ and $C_{max_k}$, $k=R,T$, which are typically
assumed to be lognormally distributed (see \cite{lacey1997}).
We denote by  $\sigma_P^2:=\big(\tfrac{1}{N_R}+\tfrac{1}{N_T}\big)\sigma^2$
the pooled variance  and by  $d:=\mu_T-\mu_R$ the difference between the expectations of the reference and the treatment group.
This yields for the difference of the means
\begin{equation}\label{normality_diff}\bar{X}_T-\bar{X}_R\sim\mathcal{N}(d,\sigma_P^2).\end{equation}
The unknown   variance $\sigma_P^2$  is estimated by
\begin{equation}\label{varest}\hat\sigma_P^{2}:=\big(\tfrac{1}{N_T}+\tfrac{1}{N_R}\big)\hat\sigma^2,
\end{equation}
where
$$\hat\sigma^2=\frac{1}{N_T+N_R-2}\sum_{k\in\{R,T\}}\sum_{i=1}^{N_k}\left(X_{k,i}-\bar{X}_k\right)^2.$$
Consequently the null hypothesis in  \eqref{hypotheses} is rejected if
\begin{equation}
\label{rej_rule_tost} \frac{\bar{X}_T-\bar{X}_R-(-\delta)}{\hat\sigma_P} \geq t_{N-2,1-\alpha}   \text{ and }  \frac{\bar{X}_T-\bar{X}_R-\delta}{\hat\sigma_P}\leq-t_{N-2,1-\alpha},
\end{equation}
where $t_{N,1-\alpha}$ is the $(1-\alpha)$-quantile of the $t$-distribution with $N-2=N_R+N_T-2$ degrees of freedom (see for example \cite{chowliu1992}).
This method is equivalent to constructing a $(1-2\alpha)$-confidence interval for $\mu_T-\mu_R$ and concluding bioequivalence if its completely contained in the equivalence interval $\left[ -\delta,\delta\right]$ (see \cite{schuirmann1987}).

The approach presented above has been extended for  model-based  bioequivalence inference by \cite{dubois2011} and will be explained in detail in Section \ref{sec42}.

\subsection{An efficient alternative to TOST}
\label{sec23}

In this section we will review  an alternative test which  is (asymptotically) the 
most powerful test in this setting. In the case of known variances this property is well known in the literature on 
mathematical statistics (see, for example, \cite{romano2005}), 
and, for the sake of completeness, a proof of optimality will be given in the Appendix \ref{sec32}, where we also review some aspects of the power of the TOST.
These  considerations motivate  the model-based method, which we will propose  in  the following Section \ref{sec43}. 

To be precise, let $ \mathcal{N}_F(d,\sigma_P^2)$ denote the folded normal distribution with parameters $(d,\sigma_P^2)$, that is
the distribution of the random variable  $|Z|$, where  $Z\sim \mathcal{N} (d,\sigma_P^2)  $.
Due to \eqref{normality_diff} we have for the absolute difference
$$\left|\bar{X}_T-\bar{X}_R\right|\sim\mathcal{N}_F(d,\sigma_P^2).$$
This result motivates the choice
of the quantile determining the decision rule of the test, which  is described in the following algorithm.

 \begin{algorithm}\label{alg2}{\rm (A more powerful test)
 		\begin{enumerate}
 			\item Estimate the parameters of interest $\hat \mu_R$ and $\hat \mu_T$
 			by $\bar{X}_R$ and $\bar{X}_T$ (for instance by non-compartmental analysis) and
 			 estimate the variance of the difference  $\bar{X}_T- \bar{X}_R$  by
			 the statistic defined in \eqref{varest}.
 			\item  Reject the null hypothesis, whenever
\be
\label{rej_rule}
\left|\bar{X}_T-\bar{X}_R\right|\ < \hat  u_{\alpha},
\ee
where 		
 $\hat  u_{\alpha}$ is the  $\alpha$-quantile of the    folded normal distribution	
 ${\cal N}_{F}(\delta,\hat\sigma_P^2)$.
 	\end{enumerate}
 	}
 	\end{algorithm}
The quantile $\hat  u_{\alpha}$ can be calculated solving the equation
$$
   \alpha = \Phi\left(\tfrac{1}{\hat  \sigma_P}(u-\delta)\right)-\Phi\left(\tfrac{1}{\hat \sigma_P}(-u-\delta)\right).
$$
Alternatively, it can directly be obtained by using statistical software, as for example the $VGAM$ package by \cite{vgam} in R.\\
The approach presented in Algorithm \ref{alg2} is extended in  Algorithm \ref{alg3}  for  model-based  bioequivalence inference, where we will estimate the parameters of interest $\hat \mu_R$ and $\hat \mu_T$  by fitting
a nonlinear mixed model to the data.

\subsection{Noncompartmental analysis}

If we are testing for PK bioequivalence considering the hypotheses in \eqref{hyp_BE} we need to calculate estimates of $AUC$ and $C_{\max}$ directly from the data.
In this regard the classical approach is given by NCA, as described for example in \cite{gabrielsson2001}. More precisely, $C_{\max}$ is directly obtained from the data, whereas $AUC$ is approximated by the linear trapezoidal rule. This means that the total area under the curve is obtained by separating it into several smaller trapezoids and summing up these areas. Of course the accuracy of this approach strongly depends on the number of measurements as this gives the number of trapezoids but it does not require a model assumption and is widely applicable.
As these methods do not take  the profile of the  blood concentration-time curve into account, we call them  NCA-based methods throughout this paper 
and they will be discussed in more detail in Section \ref{sec4}.

\section{Model-based Bioequivalence Tests}\label{sec4}
\def\theequation{4.\arabic{equation}}
\setcounter{equation}{0}

Classical NCA-based tests are a useful tool to establish bioequivalence
if  the  blood concentration profile variables  $AUC$ and $C_{\max}$ can be calculated with a reasonable precision without using
information about the form of the concentration profiles. For this purpose one usually needs a relatively dense design to determine
the area under the  curve or the maximum of the profile.  However, there are many situations, where only a sparse design is available (for some examples see \cite{hu2004}) and the NCA-based  calculation of  $AUC$ and $C_{\max}$
might be misleading as the estimates are biased in this case (see \cite{dubois2011}).  In such situations
where NCA is not reliable a model-based approach as proposed for the TOST by \cite{panhard2005} and \cite{dubois2011} might have important advantages.

Roughly speaking they  proposed to use non-linear mixed effects models (NLMEM) to describe  the blood concentration profile
and derive $AUC$ and $C_{\max}$ estimates. These quantities are then further analyzed using the methodology introduced in Section \ref{sec2}. By this approach they were able to increase the accuracy of  bioequivalence tests in the case of sparse designs.

We will use the same methodology to extend the approach presented in Section \ref{sec21} to situations with sparse designs. This new test achieves more power and simultaneously controls the type I error.

\subsection{Nonlinear mixed effects models (NLMEM)}\label{sec41}

We first consider crossover trials with $K$ periods and $N$ subjects, investigating the difference between a test and a reference treatment. A classical situation is given by the (balanced) two-period, two-sequence crossover design ($K=2$), where the $N/2$ patients receive treatment $R$ in the first period and treatment $T$ in the second one while the other $N/2$ patients
receive the treatments in the reverse order.
\\
For each subject concentrations of the drug are measured in all periods and at different sampling points.
In order to represent the dependence of the  concentration  on time for one subject we follow  \cite{dubois2011} and use a non-linear function, say  $f$ in order to fit one global model to the data, that is
\begin{equation}\label{NLMEM}
y_{i,j,k}=f(t_{i,j,k},\psi_{i,k})+g(t_{i,j,k},\psi_{i,k})\epsilon_{i,j,k},
\end{equation}
where $y_{i,j,k}$ denotes the concentration of the $i$-th subject ($i=1,\ldots N$) at sampling time $t_{i,j,k}$ ($j=1,\ldots,n_{i,k}$) of period $k$ ($k=1,\ldots K$).
In \eqref{NLMEM}  the residual errors  $\epsilon_{i,j,k}$ are independent  and standard-normally distributed random variables and the function $g$ is used to model heteroscedasticity. In particular we consider a combined error model with
\be\label{res_err} g(t_{i,j,k},\psi_{i,k})=a+b\cdot f(t_{i,j,k},\psi_{i,k}),\ee
 where the parameters $a,b\in\mathbb{R}_{\geq 0}$ account for the additive and the proportional part of the error respectively. This gives for the variance  of the  errors in  \eqref{NLMEM}
$$ \mbox{Var} ( y_{i,j,k} )
=(g(t_{i,j,k},\psi_{i,k}))^2=  | a+b\cdot f(t_{i,j,k},\psi_{i,k}) | ^2 .$$
The individual parameters $\psi_{i,k} =(\psi_{i,k,1},\ldots , \psi_{i,k,p})^\top$ (of length $p$) are defined by
\begin{equation}\label{ind_psi}
\log(\psi_{i,k,l})=\log{\lambda_l}+\beta^{T}_l Tr_{i,k}+\beta^{P}_l P_{k}+\beta^{S}_l S_{i}+\eta_{i,l}+\kappa_{i,k,l},\ l=1,\ldots,p,
\end{equation}
where $\lambda= (\lambda_{1}, \ldots , \lambda_{p})^{\top }$ denotes a vector of fixed effects, $Tr_{i,k}$, $P_k$ and $S_i$ the (known) vectors of treatment, period and sequence covariates respectively and $\beta^T$, $\beta^P$ and $\beta^S$ the vectors of coefficients of treatment, period and sequence effects. In order to account for the variability between individuals, denoted as between-subject-variability (BSV), and the variability of one subject between two periods respectively, that is the within-subject-variability (WSV), we introduce random effects  $\eta_{i}= (\eta_{i,1}, \ldots , \eta_{i,p})^{\top}$ and $\kappa_{i,k} = (\kappa_{i,k,1}, , \ldots , \kappa_{i,k,p})^{\top} $. More precisely, the random effect $ \eta_{i}$ represents the BSV of subject $i$ and $\kappa_{i,k}$ the WSV of subject $i$ at period $k$ respectively.
Throughout this section we
assume that  the random effects are normal distributed, that is
\begin{equation}\label{dis_psi}
\eta_{i}\sim \mathcal{N}(0,\Omega),\ \kappa_{i,k}\sim \mathcal{N}(0,\Gamma),\ i=1,\ldots N,\ k=1,\ldots K,
\end{equation}
with $p\times p$-dimensional covariance matrices $\Omega$ and $\Gamma$ and denote the diagonal elements of these matrices by $\omega_l^2$ and $\gamma_\ell^2$, respectively.
Finally, the vector of all  parameters in model  \eqref{NLMEM}  is given by
\be\label{theta}
\theta=(\lambda,\beta^T,\beta^S,\beta^P,\Omega,\Gamma,a,b).
\ee

For biologics with a long half-life, such as monoclonal antibodies, a parallel group design, that is each individual receives only the test or the reference treatment, may be necessary (\cite{dubois2012}). In that case, we consider only one period and the WSV can be omitted and \eqref{ind_psi} simplifies to
\begin{equation}\label{ind_psi2}
\log(\psi_{i,l})=\log{\lambda_l}+\beta^{T}_l Tr_{i}+\eta_{i,l},\ l=1,\ldots,p.
\end{equation}
Note that in this case we do not assume any period or sequence effects and hence the vector in \eqref{theta}  simplifies to $\theta=(\lambda,\beta^T,\Omega,a,b)$.
For the sake of simplicity we now introduce a vector $\beta$ which is defined by
$\beta:=\beta^{T}$ in case of parallel designs and $\beta:=(\beta^{T},\beta^S,\beta^P)$ for crossover designs. Consequently we can write for the vector of all  parameters in model  \eqref{NLMEM} $\theta=(\lambda,\beta,\Omega,\Gamma,a,b)$, where $\Gamma$ disappears in case of parallel design.

Considering now the hypotheses in \eqref{hyp_BE}
the treatment effects $\beta_{AUC}^T$ and $\beta_{C_{\max} }^T$ on $AUC$ and $C_{\max} $ respectively can be directly obtained from the parameters of the
global NLMEM. In other words, there exist functions,  $h_{\rm AUC}$, $h_{C_{\max}}$, such that
\begin{equation}\label{sec_param}
\beta_{AUC}^T=h_{\rm AUC}(\lambda,\beta),\  \beta_{C_{\max} }^T=h_{C_{\max}}(\lambda,\beta).
\end{equation}
By this we obtain an estimate for its variance using the delta method (\cite{oehlert1992}), which has been proposed by \cite{panhard2007}.
With these notations the hypotheses in \eqref{hypotheses} can be rewritten as
\begin{equation}  \label{hypneu}
H_{0}:  | \beta^T | \geq \delta \text{~ versus } ~~ H_{1}:  | \beta^T | < \delta
\end{equation}
where we do the same for $ AUC$ and $ C_{\max}$.


\subsection{Model-based TOST}
\label{sec42}
A model-based version introduced by \cite{panhard2005,panhard2007} and \cite{dubois2011} of the TOST for bioequivalence
can be obtained by fitting  the  NLMEM \eqref{NLMEM}  to the data and calculate the estimate
$\hat\beta_c^T$ of the treatment effect    $\beta_c^T$, $c=AUC,\ C_{max}$.
We can assume from the theory of mixed effects modeling (see for example \cite{demidenko2013mixed}) that this estimate $\hat\beta_c^T$ is asymptotically normal distributed and following the discussion in Section \ref{sec21}
the null hypothesis in \eqref{hypneu} is rejected whenever
\begin{equation}\label{rej_rule_tost2}
\frac{\hat\beta_c^T-(-\delta)}{SE(\hat\beta_c^T)} \geq z_{1-\alpha} \text{~~ and ~~~}
\frac{\hat\beta_c^T-\delta}{SE(\hat\beta_c^T)}\leq-z_{1-\alpha},\ c=AUC,C_{max},
\end{equation}
where $z_{1-\alpha}$ is the $(1-\alpha)$-quantile of the standard normal distribution and $SE(\hat\beta_c^T)$ is an estimate of the standard error of the estimate $\hat\beta_c^T$. \\
We obtain $SE(\hat\beta_c^T)$ by using an asymptotic approximation based on the estimated covariance matrix of the fixed effects (given by a submatrix of the inverse of the Fisher information matrix) and the Delta-method (see \cite{oehlert1992} and \cite{dubois2011} for the concrete calculation).
More precisely, considering \eqref{sec_param} and denoting the estimated covariance matrix of the fixed effects by $\hat V$, we have
\begin{equation}\label{standard_error}
SE(\hat \beta_c^T)=\sqrt{\nabla h_c(\hat \lambda,\hat\beta)\cdot\hat V\cdot\nabla h_c(\hat \lambda,\hat\beta)},~~ c=AUC, ~ C_{\max},
\end{equation}
where $\nabla h_c$  denotes the gradient of the function $h_{c}$, expressing $\beta_c^T$ as a function of the model parameters ($c=AUC$  or  $C_{\max} $).
As the functions  $h_{\rm AUC}$ and $h_{C_{\max}}$ are known, all quantities of the rejection rule given in \eqref{rej_rule_tost2} can be directly obtained from the
estimates of the parameters in model  \eqref{NLMEM}.

\subsection{Model-based optimal Bioequivalence Test}
\label{sec43}

In this section we extend the bioequivalence test described in Section \ref{sec23} to NLMEM.
It will  be shown in Section \ref{sec5}  that the new method significantly improves currently used
tests  for bioequivalence of concentration curves measured by the pharmacokinetic parameters $AUC$ and $C_{\max}$ as it can also be applied in the case of sparse designs. Further this test turns out to be more powerful than the model-based TOST described in Section \ref{sec42}, in particular for small sample sizes or data with high variability.
The adaption of Algorithm \ref{alg2} to model-based bioequivalence is very straight forward and is summarized in the following algorithm:

 \begin{algorithm}\label{alg3}{\rm (A model-based optimal bioequivalence test on $AUC$ and $C_{max}$)
		\begin{enumerate}
		\item Estimate a NLMEM to the data, resulting in the parameter estimate $\hat\theta=(\hat\lambda,\hat\beta,\hat\Omega,\hat\Gamma,\hat a,\hat b)$.  This can be done for example for parallel designs using the $saemix$ package by \cite{saemix}.
			The test statistic can be directly calculated as secondary parameter of the model parameters
			(see \eqref{sec_param}) and is given by
			\[	|\hat \beta_{c}^T|=	|h_c(\hat\lambda,\hat\beta)|,\ c=AUC,C_{max}. \]
			 Approximate the standard error of the estimate $SE(\hat\beta_c^T),\ c=AUC,C_{max},$ by using the Delta-Method as describred in \eqref{standard_error}.
			\item  Reject the null hypothesis, whenever
			\be
			\label{rej_rule_mbbot}
				|\hat \beta_{c}^T|< \hat  u_{\alpha},
			\ee
			where 		
			$\hat  u_{\alpha}$ is  the  $\alpha$-quantile of the   folded normal distribution	
		${\cal N}_{F}(\delta ,(SE(\hat\beta_c^T))^2)$.
		\end{enumerate}
	}
\end{algorithm}

Finite sample properties of this method are given in Section \ref{sec5}. 

\section{Numerical comparison of NCA- and model-based- approaches}
\label{sec5}
\def\theequation{5.\arabic{equation}}
\setcounter{equation}{0}
In this section  we investigate the finite sample properties of the different  methods by means of a simulation study.
For this purpose we  consider  eight different scenarios for parallel designs and for two-periods-two-sequence-cross-over studies respectively.
Note that the latter   represent the standard design for bioequivalence trials. More precisely, we will use the models as described in Section \ref{sec41} in order to simulate pharmacokinetic (PK) data
using a population PK model with several scenarios varying the study design, the number of sampling times per subject $n$ and the magnitude of BSV and WSV (for the cross-over designs).
The threshold for bioequivalence in \eqref{hypotheses} is as explained in Section \ref{sec2} chosen as $\delta=\log(1.25)$ in all cases under consideration.

\subsection{Settings}
\label{sec51}
We use the same PK model as described in \cite{dubois2011}, which describes concentrations ($mg/l$) of the anti-asthmatic drug theophylline, for both reference and test group. More precisely, we consider a one-compartment model with first-order absorption and first-order elimination and hence the pharmacokinetic function $f$ in \eqref{NLMEM} is defined by
\begin{equation}\label{pk_model}
f(t,D,k_a,CL/F,V/F)=\frac{F\cdot D\cdot k_a}{V(\frac{CL}{V}-k_a)}\left(\exp(-k_a\cdot t)-\exp(-\frac{CL}{V}\cdot t)\right),
\end{equation}
where $D$ is the dose, $F$ the bioavailability, $k_a$ the absorption rate constant, $CL$ the clearance of the drug, and $V$ the volume of distribution and hence $\psi$ is composed of $k_a$, $CL/F$ and $V/F$. \\
The value for the residual error model in \eqref{res_err} were set to $a=0.1$mg/l and $b=10\%$.
The dose is fixed to $D=4$mg for all subjects, and the fixed effects for the reference treatment group are $\lambda_{k_a}=1.5 \, h^{-1}$, $\lambda_{CL/F}=0.04 \, l \, h^{-1}$, and $\lambda_{V/F}=0.5 \, l$.
The  variance-covariance matrices $\Omega$ and $\Gamma$ were chosen to be diagonal and
we  investigate two different levels of variability for  the parallel  and  crossover design as specified in Table \ref{tab_var}.
To evaluate the type I error of the approaches, we simulate a treatment effect on parameters $V$ and $CL$ given by $\beta_V^T=\beta_{CL}^T=\log(1.25)$, which affects the $AUC$ and $C_{max}$ similarly, that is $\left|\mu_T-\mu_R\right|=\beta_{AUC}^T=\left|\log(AUC_T)-\log(AUC_R)\right|=\beta_{C_{max}}^T=\left|\log(C_{max_T})-\log(C_{max_R})\right|=\log(1.25).$ 
The power of the bioequivalence test will be evaluated for $\beta_{CL}=\beta_V=\log(1)$.
We will study two sampling time designs
\begin{itemize}
	\item[-] Rich design: $N=40$, $n=10$ samples taken at times $t=(0.25, 0.5, 1, 2, 3.5, 5, 7, 9, 12, 24)$ hours after dosing,
	\item[-] Sparse design: $N=40$, $n=3$ samples taken at times $t=(0.25,3.35,24)$ hours after dosing
\end{itemize}
as  described in \cite{dubois2011},  where all subjects have the same vector of sampling times.
Note that in this situation the sparse design reflects the most critical case as three sampling points are the minimum required for estimating a model with three parameters given in \eqref{pk_model}.
For each scenario, we simulate $500$ data sets. For the estimation of the model  parameters we use the SAEM algorithm (see \cite{kuhn2005maximum}). More precisely, in case of parallel designs, we use the R package $saemix$ developed by \cite{saemix} with $10$ chains and  $(300,100)$ iterations. For crossover studies, we used Monolix 2018 R2 developed  by \cite{team2018monolix} to fit the model to the data with the same number of chains and interations as for parallel designs.\\
For the standard NCA analysis (see for example \cite{gabrielsson2001}) we used the R package $MESS$ developed by \cite{mess}. As this technique is not appropriate for sparse samples we only report results   for  NCA-based methods based on rich design.



We start considering a parallel design. Two-treatments parallel trials are simulated, that is $20$ subjects receive the reference treatment R and the other $20$ subjects are allocated to the test treatment T.
Illustrations of the simulated concentrations in groups R and T under $H_0$ and $H_1$ in \eqref{hypneu} are presented in Figure $1$.
Secondly, we observe a two-periods two-sequences crossover design.
For each trial, the $20$ subjects allocated to the first sequence receive the reference treatment first and then the test treatment. The other $20$ subjects allocated to the second sequence receive treatments in the reverse order.
Table \ref{tab_var} displays all variabilities under consideration.

\begin{table}
\begin{center}
	\begin{small}
		\begin{tabular}{|c|c|c|c|c|c|c|c|c|c|}
			\hline
			\multirow{1}{*}{$\text{Design }$ }& \multirow{1}{*}{$\text{Variability Scenario }$ } & $\omega_{k_a}$ & $\omega_{V/F}$ & $\omega_{CL/F}$ & $\gamma_{k_a}$ & $\gamma_{V/F}$ & $\gamma_{CL/F}$  \\
		\hline
			\multirow{2}{5cm}{Parallel} &	Low BSV & 22 &  11 & 22 & NA & NA & NA  \\  \cline{2-8}
			&High BSV & \multicolumn{3}{c|}{52} & NA & NA & NA  \\
			\hline
		
			\multirow{2}{5cm}{Crossover } & 	Low  & $20$ &  $10$ & $20$ & $10$ & $5$ & $10$  \\   \cline{2-8}
			&High  & \multicolumn{3}{c|}{50} & \multicolumn{3}{c|}{15} \\
					\hline
		\end{tabular}\vspace{0.3cm}\\
	\end{small}
\end{center}
\caption{\it
Simulated values for the parallel and crossover design, low and high variability settings.  $\omega$ and $\gamma$ are expressed as coefficient of variation in \%. Entries "NA" correspond to "not applicable".}
\label{tab_var}
\end{table}

\begin{figure}
\begin{center}
\includegraphics[width=0.9\textwidth]{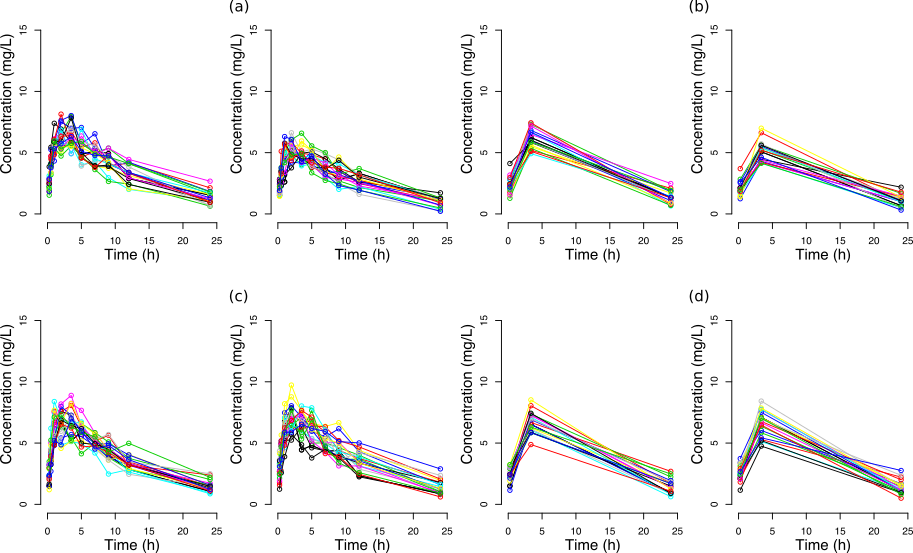}
\caption{\it Spaghetti plots of simulated concentrations for parallel design with $N=40/n=10$ ((a) and (c))) and $N=40/n=3$ ((b) and (d)), low variability under $H_0$ (top line), that is $\beta^T=\log(1.25)$ and $H_1$ (bottom line), that is $\beta^T=\log(1)$. On each plot, profiles on the left correspond to the reference group (R) and profiles on the right correspond to the treatment group (T).}\end{center}
\label{data_parallel}\end{figure}

\subsection{Results}\label{sec52}
\subsubsection{Type I error}

In Table \ref{Type_1_error}  we  show the results for all tests proposed in Sections \ref{sec2} and \ref{sec4}.
For parallel designs it becomes obvious that both the NCA-based and the model-based TOST are conservative in  settings
with a high variability, while the new approach yields a very accurate  approximation of the level. This corresponds to the empirical  findings in Section \ref{sec2}
and the theoretical arguments given in the Appendix. However, we observe a  slightly increased type I error for the sparse design
with low variability for both model-based methods, probably due to standard error underestimation as mentioned by \cite{dubois2011}.   
For rich samples and low variability all four tests  under consideration perform well
and yield an accurate approximation of the nominal level at boundary of the hypotheses, that is $\delta = \log (1.25)$.

In the  case of crossover designs the approximation of the level is very precise for all four  tests under consideration, even in the case of  high variability. This can be explained by the fact that each individual receives a test and a reference treatment and hence we have twice as much data as for the parallel designs data. However, there is a slight type I error inflation ($0.078$) for the model-based TOST considering a sparse design with high variability.
Concluding, the type I error rates are close to $\alpha$ in almost all scenarios under consideration. For increasing variances both versions of the TOST become very conservative whereas the new approach approximates the level still very precisely.

\begin{table}[h]
\begin{center}
    \begin{small}
    \begin{tabular}{|c|c|c|c|c|c|c|c|c|c|}
		\hline
\multicolumn{2}{|c}{Study Design} & \multicolumn{4}{|c|}{Parallel} & \multicolumn{4}{|c|}{Crossover} \\ \hline
\multicolumn{2}{|c|}{Sampling time} & \multicolumn{2}{|c|}{Rich}& \multicolumn{2}{|c|}{Sparse} &\multicolumn{2}{|c|}{Rich}& \multicolumn{2}{|c|}{Sparse} \\ \hline
\multicolumn{2}{|c|}{Variability} & Low & High & Low & High &  Low & High & Low & High \\ \hline
\multirow{2}{*}{NCA-TOST}& AUC & 0.052 &	\textbf{0.022} & - & -	& 0.046 & 0.042 & - & - \\
                             & $C_{\max} $	& 0.062 & \textbf{0.012} &	- & - & 0.062	&	0.070 & - & - \\ \hline
\multirow{2}{*}{NCA-BOT} & AUC & 0.052 &  0.054& - & -	& 0.046 & 0.042 & - &-\\
& $C_{\max} $	& 0.062 & 0.052&	- & -&0.062	 &	0.070 &- & -\\ \hline
\multirow{2}{*}{MB-TOST}& AUC & 0.056 &	\textbf{0.004} & \textbf{0.076} & \textbf{0.006}	& 0.056 & 0.042 & 0.038 &0.050 \\
                & $C_{\max} $	& 0.058 & \textbf{0.008} &	0.066 & \textbf{0.002} &	0.064 &	0.070 & 0.044 & \textbf{0.078} \\ \hline
\multirow{2}{*}{MB-BOT}& AUC & 0.056 &	0.064 & \textbf{0.076} & 0.034	& 0.056	 & 0.044 & 0.038 & 0.056 \\
                & $C_{\max} $	& 0.070 & 0.060 &0.070 & 0.058 &	0.064 &	0.054 & 0.044 & 0.056 \\ \hline
    \end{tabular}\vspace{0.3cm}
    \end{small}
   \end{center}
    \caption{\it Simulated type I errors of the four tests under $H_0$, where BOT denotes the test derived in Algorithm \ref{alg3}. The numbers in boldface  indicate that  the  type I error falls  outside of
     the 95\% prediction interval  $[0.0326;0.0729]$   centered at  $0.05$.}
    \label{Type_1_error}
\end{table}

\subsubsection{Power}
\label{sec53}
In order to investigate the power of the proposed methods we consider the scenarios summarized in Table \ref{tab_var}
with a treatment effect of $\beta_{AUC}^T=0$ and  $\beta^T_{C_{max}} =0$. In
Table \ref{Power} we  display the results for the four tests under consideration.
In the case of  parallel designs  we observe   that a sparse design does not affect the performance of the tests as much as the level of variability, which when high leads to a huge  loss of power for all methods.
Although in these settings the power is only close to $0.15$ for the new model-based approach, a noticeable improvement compared to the model-based TOST is visible, as
for this test the power is practically zero. For low variability the model-based tests  perform very similarly, which confirms  again
the empirical  findings in Section \ref{sec2} and some theoretical explanation for these observations is given in the appendix.
When considering rich designs the NCA-based methods achieve more power than the model-based ones but the difference turns out to be quite small. However, for sparse designs NCA-based methods are not applicable and in case of low variability we obtain a very high power for both model-based approaches.
For the cross-over designs all tests under consideration yield a power of one, irrespective of the sampling time, design and variability. This effect can again be explained by the
larger sample size and each individual receiving both treatments.

\begin{table}[h]
\begin{center}
    \begin{small}
    \begin{tabular}{|c|c|c|c|c|c|c|c|c|c|}
		\hline
\multicolumn{2}{|c}{Study Design} & \multicolumn{4}{|c|}{Parallel Design} & \multicolumn{4}{|c|}{Crossover Design} \\ \hline
\multicolumn{2}{|c|}{Sampling Time} & \multicolumn{2}{|c|}{Rich}& \multicolumn{2}{|c|}{Sparse} &\multicolumn{2}{|c|}{Rich}& \multicolumn{2}{|c|}{Sparse} \\ \hline
\multicolumn{2}{|c|}{Variability} & Low & High & Low & High &  Low & High & Low & High \\ \hline
\multirow{2}{*}{NCA-TOST}& AUC & 0.998 &	0.132 & - & -	& 1.000& 1.000 & - & - \\
& $C_{\max} $	& 0.998 & 0.056 &	- & - &	1.000 &	1.000 & - & - \\ \hline
\multirow{2}{*}{NCA-BOT} & AUC & 0.998 &  0.228& - & -	& 1.000 & 1.000 & - &-\\
& $C_{\max} $	& 0.998 & 0.154&	- & -&	1.000 &	1.000 &- & -\\ \hline
\multirow{2}{*}{MB-TOST}& AUC & 0.830 &	0.008 & 0.804 & 0.004	&1.000  & 1.000 & 1.000 & 0.998 \\
                & $C_{\max} $	& 1.000 & 0.024 &	1.000 & 0.016 &	1.000 &	1.000 &1.000  & 1.000 \\ \hline
\multirow{2}{*}{MB-BOT}& AUC & 0.838 &	0.140 & 0.808 & 0.132	&	1.000 & 1.000 & 1.000 & 1.000 \\
                & $C_{\max} $	& 1.000 & 0.138 & 1.000 & 0.116 &	1.000 &	1.000 & 1.000 & 1.000 \\ \hline
    \end{tabular}\vspace{0.3cm}
    \end{small}
    \end{center}
    \caption{\small \it Simulated power of the four tests under $H_1$,  where BOT denotes the test derived in Algorithm \ref{alg3}. }
    \label{Power}
\end{table}

\section{Conclusions}
\label{sec:conc}
In this paper we addressed the problem of sparse designs and high variability in bioequivalence studies.
As described by \cite{phillips1990} and  \cite{tsai2014} we demonstrated that in general for data with high variability methods based on the TOST suffer from a lack of power.
To address this problem  we introduced a new method using quantiles of the folded normal distribution, which  we called  bioequivalence optimal testing in this paper.
In the  case of known variances we proved in the Appendix  that this test is uniformly most powerful in this setting and has consequently more power than the TOST.
These arguments can be transferred to general bioequivalence testing   using NCA or NLMEM  if the sample variances can be estimated with reasonable accuracy.

By  means of a simulation study we compared the new procedure to the TOST, considering them both based on NCA and NLMEM. 
We demonstrated that  bioequivalence testing based on the new approach is a more powerful alternative to the  commonly used  TOST if the $AUC$ and $C_{\max}$ are obtained by NCA. 
This superiority  is also observed if these parameters are obtained by fitting an  NLMEM, in particular for data with large variability.

\bigskip

{\bf Acknowledgements} 

This work has also been supported in part by the
Collaborative Research Center ``Statistical modeling of nonlinear
dynamic processes'' (SFB 823, Teilprojekt T1) of the German Research Foundation
(DFG) and by the Food and Drug Administration (FDA) under contract 10110C.
The authors would also like to thank Dr. Martin Fink and Dr. Frank Bretz for their helpful discussions and comments on an earlier version of this paper.

\bigskip

{\bf Disclaimer} 

This article reflects the views of the authors and should not be construed to represent FDAÆs views or policies.

\begin{appendix}
\section{Theoretical comparison of  tests for bioequivalence}
 \label{sec3}
\def\theequation{3.\arabic{equation}}
\setcounter{equation}{0}

In this section we provide some theoretical explanation, why the approach presented in Section \ref{sec23}
has  more power than the TOST.  For this purpose we now  assume that variances in the reference and treatment group are known.
In this case the quantiles of the $t$-distribution in \eqref{rej_rule_tost} can be replaced by those of a normal distribution
and the power functions of all tests can be calculated explicitly. We also note that
this assumption is very well justified, if the sample sizes in both groups are sufficiently large.
In other words:  all arguments presented
in this section can be applied to the NCA-based tests discussed in Section \ref{sec21} and \ref{sec23}  provided that
the sample sizes are sufficiently large.  A similar comment applies to the model-based test for bioequivalence introduced in Section \ref{sec4}.
We begin with a discussion of the TOST.

\subsection{The two one-sided  Test (TOST)}\label{sec31}

Consider the rejection rule of the TOST defined in \eqref{rej_rule_tost}, where we replace the estimate of the (pooled) variance $\sigma_P^2=\tfrac{\sigma^2}{N_T}+\tfrac{\sigma^2}{N_R}$ by its true value and the quantile $t_{N-2,1-\alpha}$ by the $(1-\alpha)$ quantile of the standard normal distribution denoted by $z_{1-\alpha}$.
If $z_{1-\alpha} > \delta / \sigma_P$ the probability of rejection is $0$ (because the conditions in \eqref{rej_rule_tost} are contradicting).
On the other hand, and more importantly,   if $z_{1-\alpha} \leq \delta / \sigma_P$ the  probability of rejection for the test \eqref{rej_rule_tost}   is given by
 \begin{eqnarray}\label{prob_rej_TOST}
\Psi_{\small \rm TOST}(d)&:=&\mathbb{P}_d\Big (\tfrac{\bar{X}_T-\bar{X}_R+\delta}{\sigma_P}\geq z_{1-\alpha}, \tfrac{\bar{X}_T-\bar{X}_R-\delta}{\sigma_P}\leq-z_{1-\alpha}\Big  )\nonumber\\
&=&\mathbb{P}_d\Big  (z_{1-\alpha}-\tfrac{\delta+d}{\sigma_P}\leq\tfrac{\bar{X}_T-\bar{X}_R-d}{\sigma_P} \leq-z_{1-\alpha}+\tfrac{\delta-d}{\sigma_P}
\Big ) \nonumber\\
&=&\Phi\Big  (-z_{1-\alpha}+\tfrac{\delta-d}{\sigma_P}\Big )-\Phi\Big (z_{1-\alpha}-\tfrac{\delta+d}{\sigma_P}\Big ),
\end{eqnarray}
where $\Phi$ denotes the distribution function of the standard-normal distribution.
From this formula we draw the following conclusions (if $z_{1-\alpha} \leq \delta / \sigma_P$):
\begin{itemize}
\item[(1)]  The test \eqref{rej_rule_tost} controls its level. For example, if   $d>\delta$  we have
$$\Psi_{\small \rm TOST}(d) < \Phi\left(-z_{1-\alpha}\right)=\alpha$$
and with a similar argument the same inequality can be derived   for $d<-\delta$.
\item[(2)]  At the ''boundary'' of the null hypothesis (that is $d  \in \{ - \delta, \delta \}$) we have
\begin{eqnarray*}
\Psi_{\small \rm TOST}(\pm \delta) &= \alpha-\Phi\left(z_{1-\alpha}-\tfrac{2\delta}{\sigma_P}\right)\leq\alpha,
\end{eqnarray*}
As $\Phi\left(z_{1-\alpha}-\tfrac{2\delta}{\sigma_P}\right)$  converges to $0$ if $\tfrac{\delta}{\sigma_P}$ converges to infinity, we expect that  the
  level of the test \eqref{rej_rule_tost} is close to $\alpha$  at the  ''boundary'' of the null hypothesis, if  $\sigma$ is small.
This happens, for example, if the variance $\sigma^{2}$ (and hence the pooled variance $\sigma_P^{2}$) is small or, alternatively, if the sample
sizes $N_{R}$  and $N_{T}$ in both groups are very large. On the other hand  the test \eqref{rej_rule_tost}  is conservative if the
variance $\sigma^2$  is large.
In the extreme case $\tfrac{\delta}{\sigma_P}=z_{1-\alpha}$ we have
\begin{eqnarray*}
\Psi_{\small \rm TOST}(\pm \delta) &= \alpha-\Phi\left(-z_{1-\alpha}\right)=0.
\end{eqnarray*}
\end{itemize}

\subsection{The uniformly most powerful approach}
\label{sec32}
Similar to the TOST the test proposed in Section \ref{sec23} simplifies under the additional assumption of a known variance. As the variance is assumed to be known, the null hypothesis is rejected, whenever,
			\be
			\label{rej_rule2}\left|\bar{X}_T-\bar{X}_R\right|\ <u_{\alpha},
			\ee
			where  $u_{\alpha}$ denotes  the  $\alpha$-quantile of the
 			folded normal distribution  ${\cal N}_{F}(\delta ,\sigma_P^2)$.
 The following result shows that the test defined by  \eqref{rej_rule} is the uniformly most powerful test for the hypotheses  \eqref{hypotheses}.
 It is well known in the mathematical statistics literature and we present a proof here  for the sake of completeness
 (see also \cite{lehmann2006},   \cite{romano2005} or   \cite{wellek2010testing}).

 \begin{satz}\label{thm1}
	 The test defined by  \eqref{rej_rule} is the uniformly most powerful (UMP) for the hypotheses  \eqref{hypotheses}. Moreover, among all tests
	 for the hypotheses  \eqref{hypotheses}
	 with power function $\Psi$ satisfying $\Psi (\delta) = \Psi (-\delta) = \alpha$ the test defined by  \eqref{rej_rule}
	 has also minimal type I error.
\end{satz}

{\bf Proof:}
In order to prove optimality recall \eqref{normality_diff}, that is  $X=\bar{X}_T-\bar{X}_R \sim {\cal N} (d,\sigma_P^{2})$, and note that
the hypotheses in \eqref{hypotheses} can be rewritten as
\be\label{hypotheses1}
H_0:  | d |\geq \delta \text{ vs. } H_1:  |d |< \delta ~.
\ee
The test \eqref{rej_rule} rejects  the null hypothesis whenever $ \left|\bar{X}_R-\bar{X}_T\right|\ <u_{\alpha}$,
where $u_\alpha$ is the quantile of the folded normal distribution with parameters $(	\delta,\sigma_P^2)$, which is defined by
\begin{eqnarray}\label{quantile}
\alpha&=& \mathbb{P}\left( \left| \mathcal{N}(	\delta ,\sigma_P^2)\right|  \leq u_\alpha\right)
= \Phi\left(\tfrac{1}{\sigma_P}(u_{\alpha}-\delta)\right)-\Phi\left(\tfrac{1}{\sigma_P}(-u_{\alpha}-\delta)\right)~.
\end{eqnarray}
The probability of rejection is now given by
 \begin{eqnarray}\label{prob_rej}
 \mathbb{P}_d(\left|\bar{X}_T-\bar{X}_R\right|\ <u_{\alpha})
 &=&\mathbb{P}_d(-u_{\alpha}<\bar{X}_T-\bar{X}_R <u_{\alpha}) \nonumber\\
 &=&\mathbb{P}_d\left(\tfrac{1}{\sigma_P}(-u_{\alpha} -d)<\tfrac{\bar{X}_T-\bar{X}_R -d}{\sigma_P} <\tfrac{1}{\sigma_P}(u_{\alpha}-d)\right) \nonumber\\
 &=&\Phi\left(\tfrac{1}{\sigma_P}(u_{\alpha}-d)\right)-\Phi\left(\tfrac{1}{\sigma_P}(-u_{\alpha}-d)\right),
 \end{eqnarray}
where $\Phi$ denotes the distribution function of the standard-normal distribution.

On the other hand the uniformly most powerful test for the problem \eqref{hypotheses1} is well known, see for example
 Theorem 6 in Section 3.7 of \cite{lehmann2006} or Example 1.1 in \cite{romano2005}
 This test reject the null hypothesis in \eqref{hypotheses1}, whenever
 $$
 |\bar{X}_T-\bar{X}_R | < C
 $$
 where the constant $C=C( \alpha, \delta, \sigma_P) $ is the unique solution of the equation
   \begin{eqnarray}  \label{rom}
   \alpha&=&\Phi\left(\tfrac{1}{\sigma_P}(C-\delta)\right)-\Phi\left(\tfrac{1}{\sigma_P}(-C-\delta)\right)
 \end{eqnarray}
 [see Example 1.1 in \cite{romano2005}].
 As the equations \eqref{quantile}  and \eqref{rom} coincide, it follows that $u_{\alpha} = C$ and the test
 \eqref{rej_rule} coincides with the UMP test for the hypotheses \eqref{hypotheses}.
 \hfill $\Box$

\bigskip

\begin{figure}[h]
	\subfigure{\includegraphics[width=0.32\textwidth]{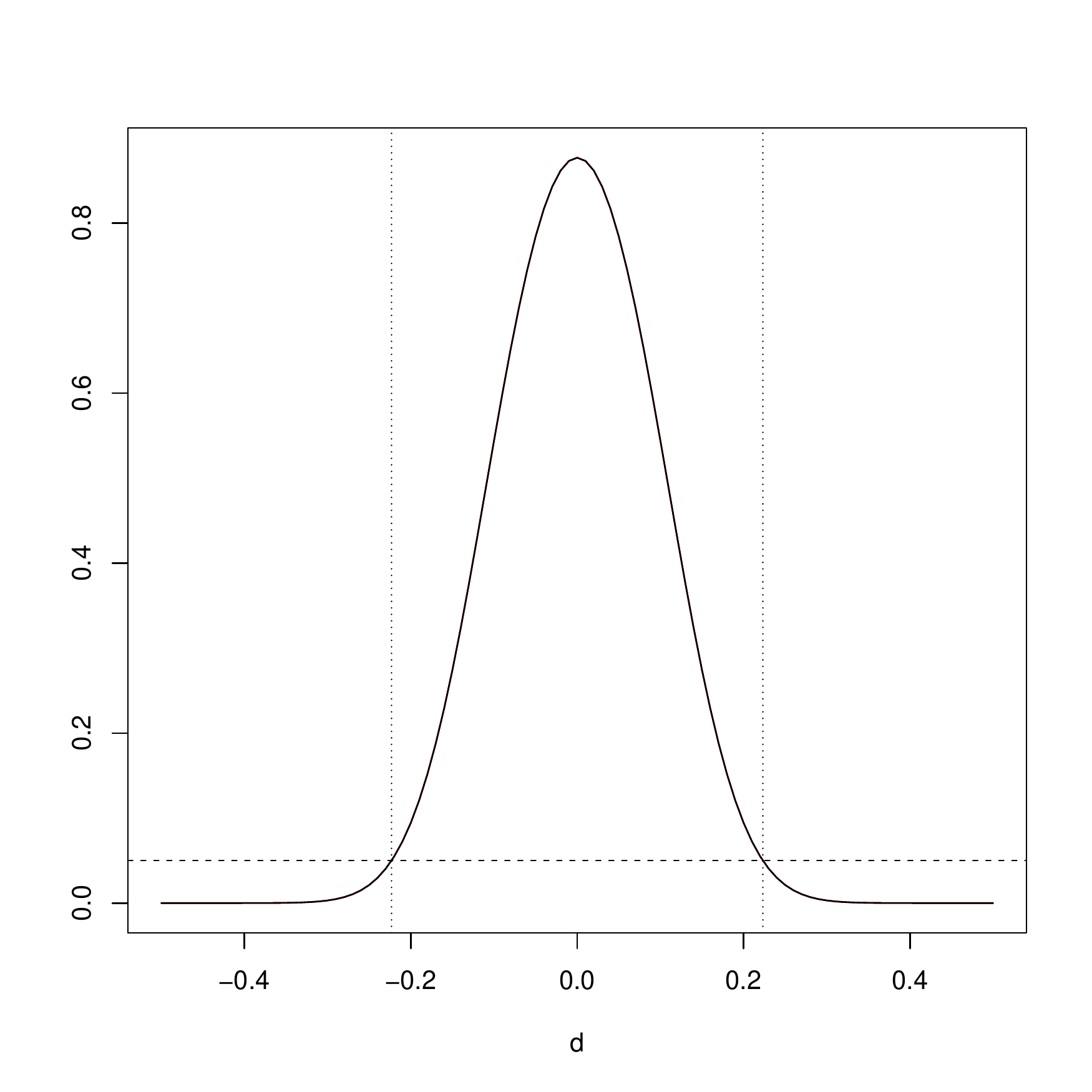}}
	\subfigure{\includegraphics[width=0.32\textwidth]{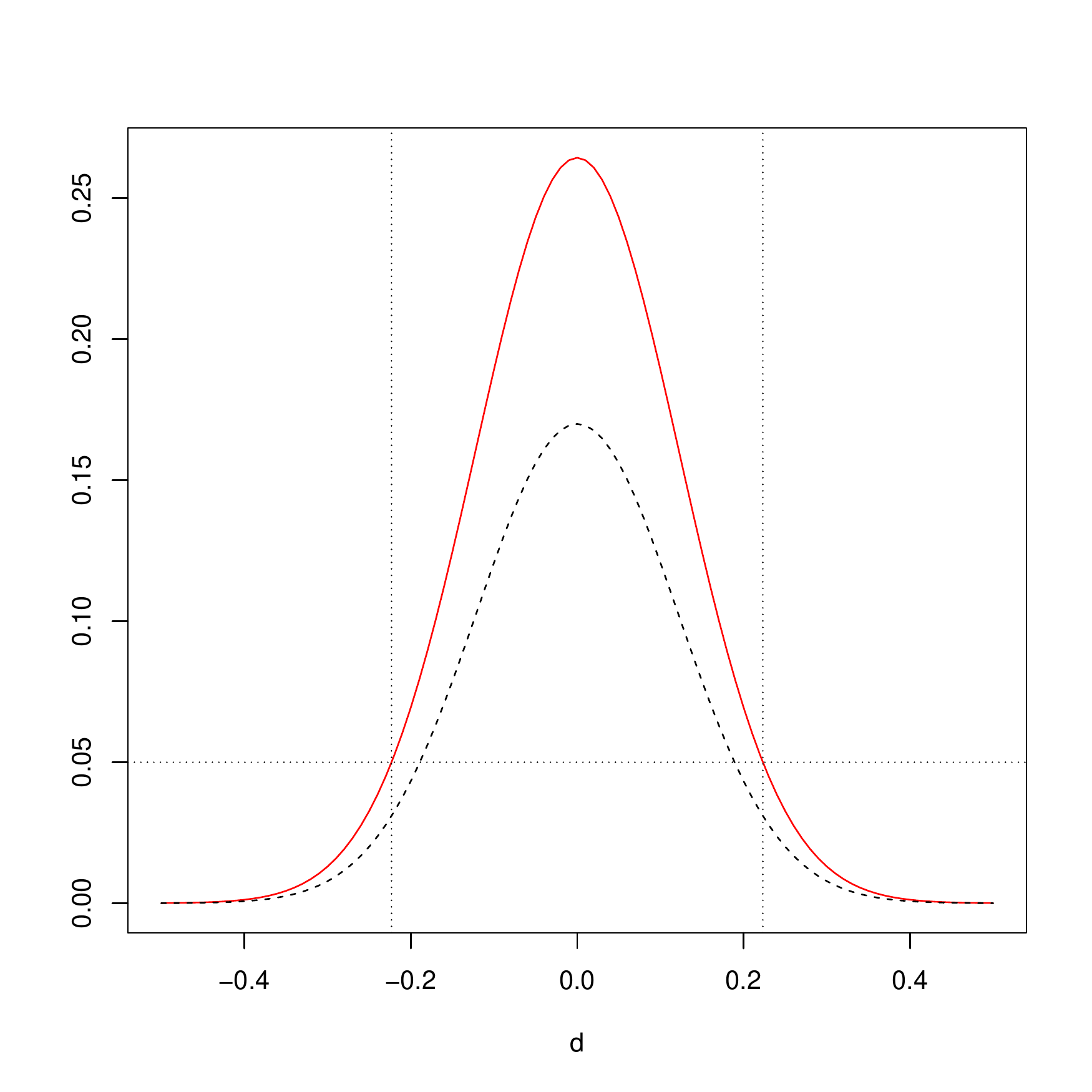}}
	\subfigure{\includegraphics[width=0.32\textwidth]{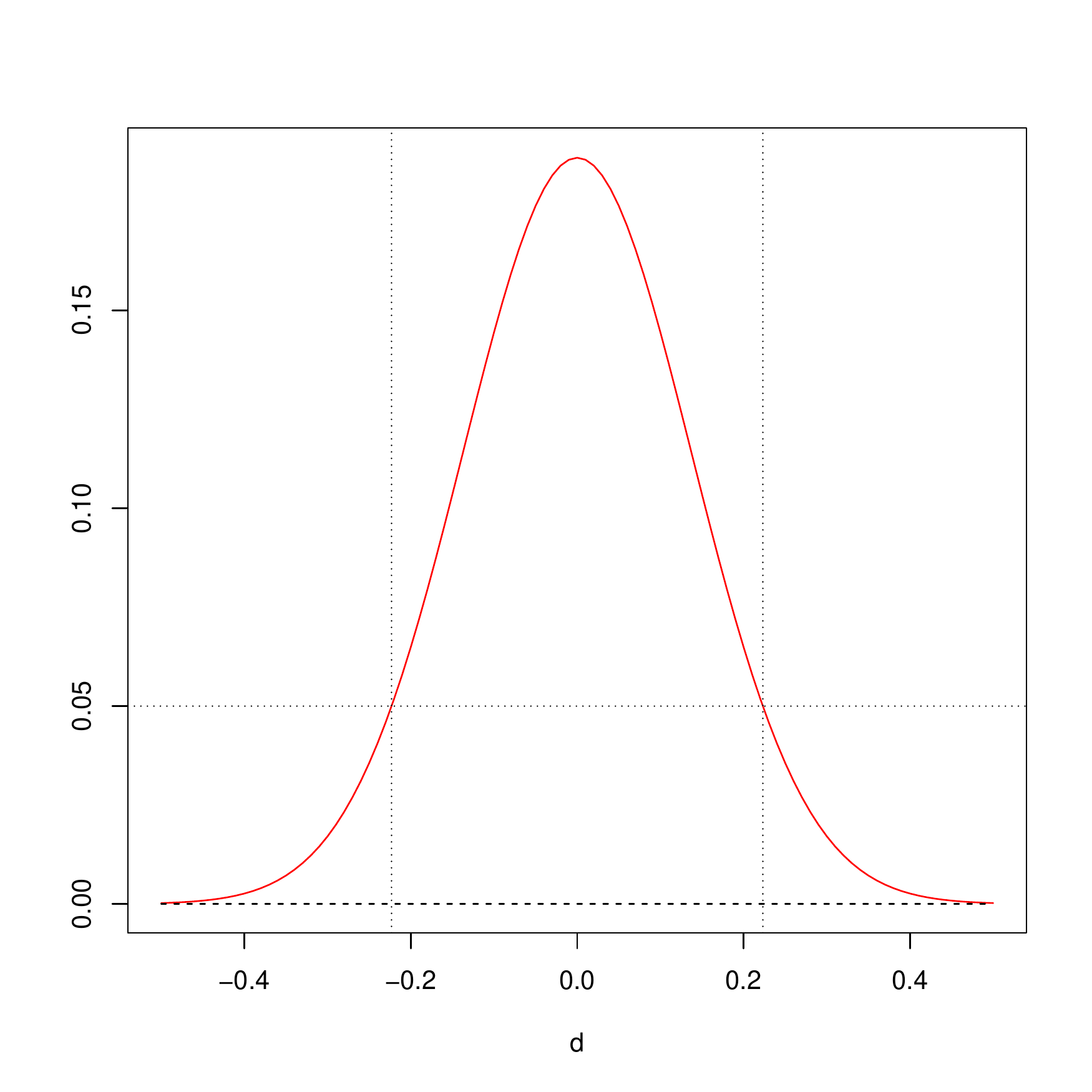}}
	\caption{{\it Power curves of the tests  \eqref{rej_rule} (solid red line) and  the test  \eqref{rej_rule_tost} (dashed line)  for different $\sigma_P=0.07$, $\sigma_P=0.12$ and $\sigma_P=\tfrac{\log(1.25)}{z_{1-\alpha}}\approx 0.14$ (from left to right). The horizontal line  indicates the level $\alpha=0.05$
	and the vertical lines mark the threshold ($\pm \delta=\pm log(1.25)$). }}
\label{fig1}\end{figure}

As a consequence of Theorem \ref{thm1} the test proposed in Section \ref{sec23}
has always more power than the test defined by  \eqref{rej_rule_tost}.  This is indicated in   Figure \ref{fig1}, where we display the power of both tests in different scenarios ($\alpha=0.05$, $\delta =\log(1.25)$).
The left panel shows the power curves for  $\sigma_P^{2}=0.0049$. In this case the curves basically coincide (although the power of the  test \eqref{rej_rule}
is slightly larger as stated in Theorem \ref{thm1}). For increasing variance ($\sigma_P^2=0.0144$) it becomes obvious that the power of the test  \eqref{rej_rule} is much higher than that for the TOST. This effect becomes even clearer in the right panel ($\sigma_P^2=\big(\tfrac{\log(1.25)}{z_{1-\alpha}}\big)^2\approx 0.14^2$), where the power curve of the TOST is identical to zero.

 \end{appendix}
\end{document}